# Software Implementation of the SNOW 3G Generator on iOS and Android Platforms

J. Molina-Gil[1], P. Caballero-Gil[1], C. Caballero-Gil[1], A. Fúster-Sabater[2]

[1]Department of Computer Engineering. University of La Laguna. Spain.
Email:{jmmolina, pcaballe, ccabgil}@ull.es
[2]Institute of Applied Physics. Spanish National Research Council. Madrid, Spain.
Email: amparo@iec.csic.es

**Abstract.** The standard for wireless communication of high-speed data in mobile phones and data terminals, called LTE (Long-Term Evolution) and marketed as 4G/LTE, is quickly being adopted worldwide. The security of this type of communication is a crucial factor mainly due to its mobile and wireless nature. This work includes a practical analysis of the SNOW 3G generator used to protect the confidentiality and integrity in LTE communications. In particular, several techniques to perform multiplications and LFSR operations have been studied and implemented on both iOS and Android platforms. The evaluation of those implementations led to some conclusions that could be used to improve the efficiency of future implementations of the standard.

## 1 Introduction

The fast substitution of 3G/UMTS by 4G/LTE technology is mainly due to the large increase of mobile data consumption and the development of applications and broadband services that are very demanding. This evolution, as was the case in each of the evolutions of previous telecommunications systems, has involved major security improvements through learning from weaknesses and attacks on their predecessors.

As for the encryption used to protect the confidentiality of mobile conversations, the evolution has been as follows. First, the stream ciphers A5/1 and A5/2 were developed for the 2G/GSM mobile phone standard but serious weaknesses were identified in both ciphers and consequently the encryption contained in the UMTS/3G standard was replaced by a completely different system, the Kasumi block cipher. In 2010, Kasumi was broken using very modest computational resources, and again the system had to be replaced by the 4G/LTE standard, where the SNOW 3G generator is used to protect confidentiality and integrity respectively through the UEA2 and UIA2 algorithms published in 2006 by the 3GPP Task Force [1].

The main goal of this work is to carry out a practical analysis of implementations of the SNOW 3G generator on the tow main mobile platforms in order to try to increase speed data rates in mobile phones and devices with limited resources.

Although the software implementation of the SNOW 3G generator done in this work cannot be considered the standard one in current mobile phones, the performance evaluation of the software implementations provided in this paper could be helpful in future developments of the standard.

This paper is organized as follows. A brief discussion of some related works is included in Section 2. Then, Section 3 presents the main concepts and notations regarding the SNOW 3G generator used in this paper. Section 4 gives some details of the implementations carried out on iOS (iPhone Operating System) and Android platforms, and their performance evaluation. Finally, Section 5 closes this paper with some conclusions and ideas for future works.

## 2 Related Work

SNOW 1.0 was the first member of the SNOW family. It was developed for the European NESSIE project whose main objective was to identify secure cryptographic primitives. However, very soon several effective attacks against SNOW 1.0 were reported. One of the first ones was a key recovery attack that requires a known output sequence of length 295 and has an expected complexity of 2224 [4]. Another one was a distinguishing attack [5], which also requires a known output sequence of length 295 and has a similar complexity.

The efficiency of the attacks launched against SNOW 1.0 showed several weaknesses in its design, so a more secure version called SNOW 2.0 was proposed, based on design principles similar to those of the stream cipher called SOSEMANUK that had been one of the final four Profile 1 (software) ciphers selected for the eSTREAM Portfolio [6]. SNOW 2.0 is nowadays one of two stream ciphers chosen for the ISO/IEC standard IS 18033-4 [7].

During its evaluation by the European Telecommunications Standards Institute (ETSI), the design of SNOW 2.0 was further modified to increase its resistance against algebraic attacks [8] with the result called SNOW 3G. A survey of this generator was given by ETSI in [9], but a full evaluation of the design of SNOW 3G has not been made public till now.

Different designers and external reviewers have showed that SNOW 3G has a remarkable resistance against linear distinguishing attacks [10, 11]. However, SNOW 3G has suffered other types of attacks. One of the earliest and simplest cryptanalytic attack attempts was the fault attack proposed in [12]. An approach to address this issue includes the use of nonlinear error detecting codes. A cache-timing attack [13] on SNOW 3G, based on empirical timing data, allows retrieving the full encryption state in seconds without any known keystream. This type of attack is based on the fact that operations such as permutations and multiplication by the constant $\alpha$ and its inverse are actually implemented using lookup tables. The work [14] describes a study of the resynchronization mechanism of SNOW 3G using multiset collision attacks, showing a simple 13-round multiset distinguisher with a complexity of 28.

The SNOW 3G generator has undergone several review works [15, 16]. This work provides a new study whose focus is on the practical insight of implementations.

The software implementations performed on iOS and Android platforms for the analysis done in this work have been developed taking as starting points the works [17] and [18].

## 3 Description of the SNOW 3G Generator

Pseudo-random generators are the main part of stream ciphers because they are used to generate the keystream sequence whose bits are bitwise XORed with the plaintext to produce the ciphertext. The main advantage of stream ciphers is that they are lightweight and can operate at a high speed, so they are very suitable for power-constrained devices such as mobile phones. On the other hand, Linear Feedback Shift Registers (LFSRs) are the main part of typical nonlinear pseudo-random generators used in most stream ciphers. That is exactly the case of the SNOW 3G generator analysed in this work.

The notation used in this paper is as follows:

| | |
|---|---|
| $GF(2)=\{0,1\}$ | Galois Field with two elements 0 and 1. |
| $GF(2)[x]$ | Ring of polynomials in the variable x with coefficients in GF(2). |
| $d$ | Degree of a polynomial. |
| $p(x)$ | Primitive polynomial of degree din GF(2)[x]. |
| $GF(2^d)$ | Extension field of GF(2) defined by p(x), with $2^d$ elements. |
| $GF(2^d)[x]$ | Ring of polynomials in the variable x with coefficients in $GF(2^d)$. |
| $\beta \in GF(2^8)$ | Root of the GF(2)[x] polynomial $x^8 + x^7 + x^5 + x^3 + 1$. |
| $\alpha \in GF(2^{32})$ | Root of the $GF(2^8)[x]$ polynomial $x^4 + \beta^{23}x^3 + \beta^{245}x^2 + \beta^{48}x + \beta^{239}$. |
| $s_t$ | 32-bit stage of an LFSR. |
| $=$ | Assignment operator. |
| $\oplus$ | Bitwise XOR operation. |
| $\boxplus$ | Integer addition modulo $2^{32}$. |
| $\|\|$ | Concatenation of two operands. |

Fig. 1 shows that the SNOW 3G generator consists of two main components: an LFSR and a Finite State Machine (FSM).

On the one hand, the LFSR has 16 stages $s_0$, $s_1$, $s_2$,..., $s_{15}$, each holding 32 bits. Its feedback is defined by a primitive polynomial over the finite field $GF(2^{32})$, and consists of two multiplications, one by a constant $\alpha \in GF(2^{32})$ and the other by its inverse, as described by the relationship: $s_{t+16} = \alpha\, s_t \oplus s_{t+2} \oplus \alpha^{-1} s_{t+11}$, for $t \geq 0$.

On the other hand, the FSM is the nonlinear component of the generator. It involves two inputs of the LFSR, which are the $s_5$ and $s_{15}$ stages contents. The FSM is based on three 32-bit registers R1, R2 and R3, and two substitution boxes S1 and S2 used to update the registers R2 and R3. Both S-Boxes S1 and S2 map each 32-bit input to a 32-bit output by applying various combinations of a basic S-box on each of the four bytes of the input. However, while the S-box S1 is based on the AES (Advanced Encryption Standard) S-box, the basic S-box S2 was specially designed for SNOW 3G. Finally, the mixing operations in the FSM are bitwise XOR operations and integer additions modulo $2^{32}$.

The LFSR of SNOW 3G has two different modes of operation, the initialisation mode and the keystream mode. On the one hand, when the initialisation is done, the generator is clocked without producing any output. On the other hand, in the keystream mode, with every clock pulse the generator produces a 32-bit word. Therefore, SNOW 3G is considered a word-oriented generator because it outputs a sequence of 32-bit words under the control of a 128-bit key and a 128-bit Initialization Vector IV.

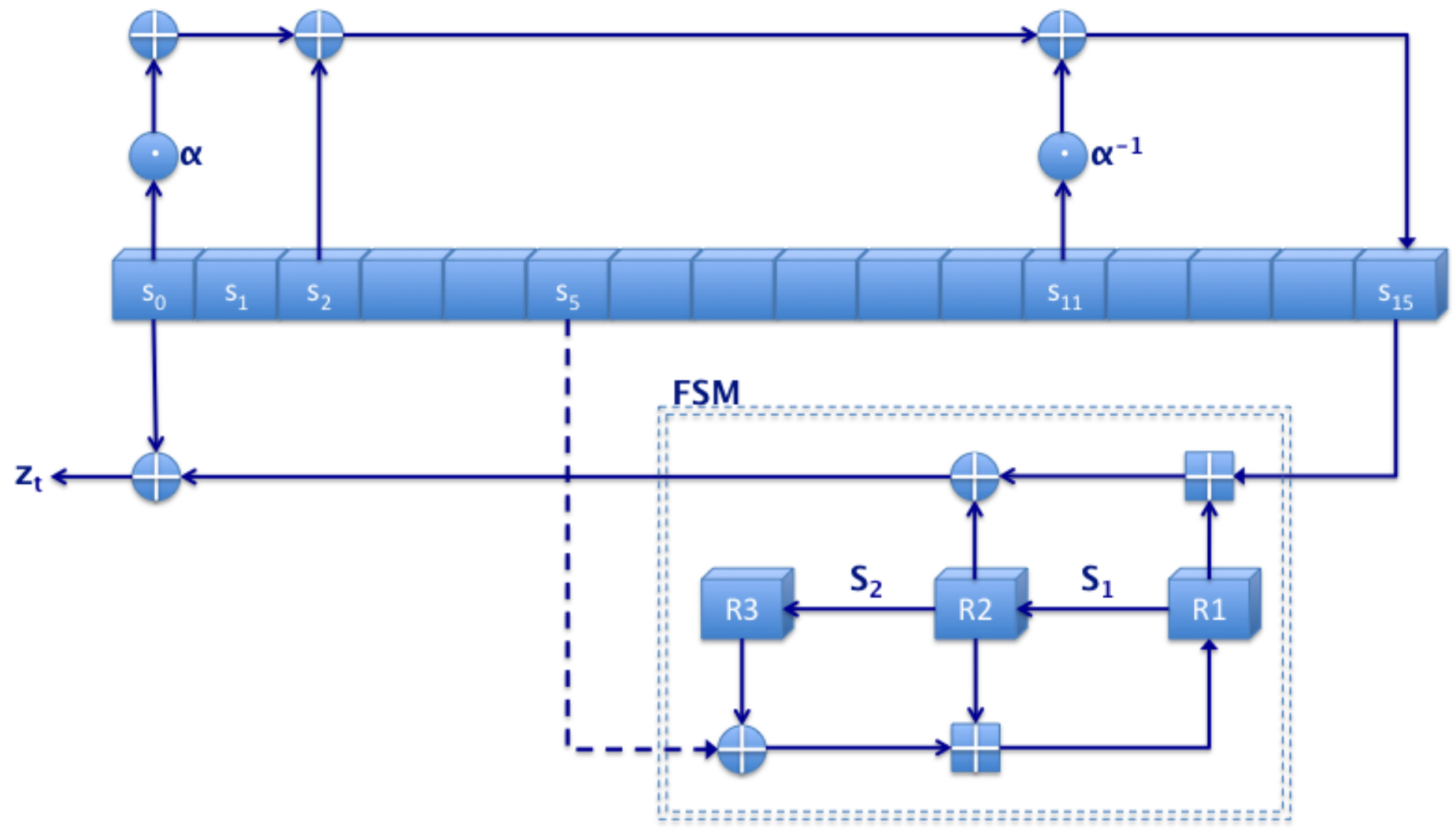

**Fig. 1.** SNOW 3G Generator

Regarding the implementation of SNOW 3G, which is the main goal of this work, several observations can be made. First, the two multiplications involved in the LFSR can be implemented as a byte shift together with an unconditional XOR with one of $2^8$ possible patterns, as shown below. In particular, as explained in [18], both multiplications can be based on lookup tables. On the one hand, since α is a root of the primitive $GF(2^8)[x]$ polynomial $x^4 + \beta^{23}x^3 + \beta^{245}x^2 + \beta^{48}x + \beta^{239}$, we can represent any element of $GF(2^{32})$ either with a polynomial in $GF(2^8)[x]$ of degree less than 4, or with a word of 4 bytes corresponding to the 4 coefficients in such a polynomial so a typical binary implementation of the multiplication of α and any 4-byte word ($c_3$, $c_2$, $c_1$, $c_0$) in $GF(2^{32})$ can be based on lookup tables ($c\beta^{23}$, $c\beta^{245}$, $c\beta^{48}$, $c\beta^{239}$), $\forall c \in GF(2^8)$. Also a typical binary implementation of the multiplication of $\alpha^{-1}$ and any 4-byte word ($c_3$, $c_2$, $c_1$, $c_0$) in $GF(2^{32})$ can be based on lookup tables ($c\beta^{16}$, $c\beta^{39}$, $c\beta^{6}$, $c\beta^{64}$), $\forall c \in GF(2^8)$. On the other hand, since β is a root of the primitive polynomial $x^8 + x^7 + x^5 + x^3 + 1$, any element of $GF(2^8)$ can be represented either with a polynomial in GF(2)[x] of degree less than 8, or with a byte whose bits correspond to the coefficients in such a polynomial so that the implementation of the multiplication of those can also be done using a modified version of the Peasant algorithm, as in AES encryption.

## 4 Implementations and Evaluation

This paper discusses software implementations of the SNOW 3G generator in the two main mobile phone platforms. On the one hand, it was implemented for the iOS platform using Objective-C as programming language. On the one hand, it was also implemented for the Android platform using Java as programming language.

These two platforms have been chosen for the study of this work for two main reasons. On the one hand, currently Android and iOS are the most commonly used mobile platforms, and on the other hand, each one of these two platforms uses a different programming language. iOS uses Objective-C, which is a compiled language, provided as a front-end for GCC like C and C++, and generates native code that can be directly executed by CPU. Android uses Java, which is a compiled, interpreted and platform independent language that does not generate any machine language code after compiling source files. In Java the real compilation into native code is done by a program called Just In Time (JIT) compiler, which is an optimization done on a Java Virtual Machine (JVM). When a JVM interprets Java byte code, it also collects interesting runtime statistics, such as the part of the code that is always being executed. Once the JVM has enough data to make a decision, JIT can compile that part of the code into native code. Then the machine will directly run this native code, without being interpreted by the JVM. For this reason, and as discussed below, the results for the implementations of the proposed mechanisms will be different for each used platform.

The first aspect that has been taken into consideration here is that LFSRs have been traditionally designed to operate over the binary Galois field GF(2). This approach is appropriate for hardware implementations, but its software efficiency is quite low. The main reason of this fact is the required content shift of each stage in each clock pulse since shift registers are structures with a very difficult software implementation due to their inherent parallel nature. In hardware implementation, the shifts of all the stages occur simultaneously, so the whole process can be performed in a single clock pulse. However, in a software implementation, this process is iterative and very costly. This negative effect is even worse if the LFSR operates over the binary Galois field GF(2) because software uses the byte as the minimum representation. Nevertheless, since microprocessors of most mobile phones have a word length of 32 bits, the LFSR implementation is expected to be more efficient for extension fields, such as $GF(2^{32})$. Thus, since SNOW 3G is defined over the extension field $GF(2^{32})$, its implementation is adequate for the architecture that supports current mobile phones. The second key aspect is related to arithmetic operations and specifically to the multiplication on extension fields of GF(2) because the feedback function in SNOW 3G involves several additions and multiplications, and the multiplication is the most computationally expensive operation.

In this section, we study and compare different software implementations in order to find the optimal one for two types of devices with limited resources. Taking into consideration that the two main mobile platforms are iOS and Android, in this work we have performed experiments on different software implementations for both platforms in order to find the optimal one for each one. In particular, we have performed several studies on an iPhone 3GS for the iOS operating system, and on an LG 9 Optimus for the Android operating system. Their main characteristics of these smartphones are described in Table 1.

**Table 1.** Devices Used for the Evaluation

| **iPhone 3GS** | | | |
|---|---|---|---|
| ***Architecture*** | ***CPU Frequency*** | ***CacheL1I/L1D/L2*** | ***RAM*** |
| Armv7-A | 600 MHz | 16 Kb/16 Kb/256 Kb | 256 MB |
| **LG 9 Optimus** | | | |
| ***Architecture*** | ***CPU Frequency*** | ***Cache L1I/L1D/L2*** | ***RAM*** |
| ARM Cortex-A9 | 1000 MHz | 64Kb//1024Kb | 1024 MB |

All the results shown in this work for the iOS platform have been obtained, as in [18], by using Instruments, which is a tool for analysis and testing of performance of OS X and iOS code. It is a flexible and powerful tool that lets you track one or more processes, and examine the collected data. On the other hand, results for Android platform have been obtained using Traceview, which is a graphical viewer for execution logs that can be used to help to debug application and profile code performance.

Test results shown in this work correspond to the average of 10 runs in which $10^7$ bytes of keystream sequence are generated using the two platforms described above. Table 2 shows the total time (in milliseconds) for every SNOW 3G function, as explained below. The evidences indicate that the multiplication (called MULxPow) is the most expensive function in both cases. The second most expensive function is the shift register (called ClockLFSRKeyStreamMode), which is performed in each clock pulse.

**Table 2.** Function Performance in Recursive Mode

| **Summary** | | | | | | |
|---|---|---|---|---|---|---|
| | **iOS** | | | **Android** | | |
| **Function** | ***Time (ms)*** | ***%*** | ***Throughput (Mbps)*** | ***Time (ms)*** | ***%*** | ***Throughput (Mbps)*** |
| MULxPow | 29054,9 | 93,7 | 3,4 | 1840,9 | 34,3 | 54,3 |
| ClockLFSRKeyStreamMode | 572 | 1,8 | 174,8 | 1473,9 | 27,5 | 67,8 |
| DIValpha | 356,6 | 1,1 | 280,4 | 349,9 | 6,5 | 285,8 |
| MULalpha | 326,8 | 1,1 | 306 | 1499,5 | 27,9 | 66,7 |
| main | 264,7 | 0,9 | 377,8 | 122,2 | 2,3 | 818,3 |
| ClockFSM | 180,3 | 0,6 | 554,6 | 42 | 0,8 | 2381 |
| S1 | 129,9 | 0,4 | 769,8 | 11,6 | 0,2 | 8620,7 |
| S2 | 128,1 | 0,4 | 780,6 | 28,3 | 0,5 | 3533,6 |
| Generator | 1,3 | 0 | 76923,1 | 0,2 | 0 | 500000 |
| **Total Time** | 31014,6 | 100 | 3,2 | 5368,5 | 100 | 18,6 |

From the obtained data, it is clear that the performance in recursive mode is not appropriate for iOS because the resulting throughput in that case is very low. On the other hand, the behaviour of this recursive implementation in Android is much better, and indeed it is better than some previous software implementations [19], where the obtained throughput in some cases was 18.3 Mbps. A fact that can explain this is that the obtained percent of total processing time for the multiplication in our recursive implementation for iOS is much higher than the one obtained in those analyses.

Below we study two different techniques to improve software implementations of multiplications and several techniques for LFSR operations proposed in [11].

### 4.1 Multiplication

Taking into account the implementation proposed in [1], Table 2 shows that the most time consuming function on both platforms is the MULxPow used to implement the multiplication by α and by $\alpha^{-1}$. Each multiplication can be implemented either as a series of recursive byte shifts plus additional XORs, or as a lookup table. In each clocking of the LFSR, the feedback polynomial uses two functions $MUL_\alpha$ and $DIV_\alpha$:

$$MUL_\alpha \quad MUL_xPOW(c,23,0xA9) \,||\, MUL_xPOW(c,245,0xA9) \,||\, MUL_xPOW(c,48,0xA9) \,||\, MUL_xPOW(c,239,0xA9)$$

$$DIV_\alpha \quad MUL_xPOW(c,16,0xA9) \,||\, MUL_xPOW(c,39,0xA9) \,||\, MUL_xPOW(c,6,0xA9) \,||\, MUL_xPOW(c,64,0xA9)$$

The first method might be more appropriate for systems with limited memory resources, because it does not require a large storage. However, as we can see in Table 2, it involves a significant computational cost.

The second method is based on lookup tables. When memory resources are not a constraint, the fastest implementation for a good multiplication performance is based on the use of lookup tables. In this proposal, this is applied when the sizes of the tables are reasonable. In SNOW 3G, the multiplication results are stored in a table containing 256 elements.

**Table 3.** Function Performance with Lookup Tables

| **Computational Cost** | | | | | | |
|---|---|---|---|---|---|---|
| | **iOS** | | | **Android** | | |
| **Function** | ***Time (ms)*** | ***%*** | ***Throughput (Mbps)*** | ***Time (ms)*** | ***%*** | ***Throughput (Mbps)*** |
| ClockLFSRKeyStreamMode | 347,3 | 29 | 287,9 | 1344 | 68,5 | 74,4 |
| main | 277,2 | 23,1 | 360,8 | 341,1 | 17,4 | 293,2 |
| ClockFSM | 182,2 | 15,2 | 548,8 | 36,6 | 1,9 | 2732,2 |
| S1 | 146 | 12,2 | 684,9 | 10,6 | 0,5 | 9434 |
| S2 | 138 | 11,5 | 724,6 | 24 | 1,2 | 4166,7 |
| GenerateKeystream | 107,3 | 8,9 | 932 | 203 | 10,4 | 492,6 |
| Generator | 1,3 | 0,1 | 76923,1 | 1,9 | 0,1 | 52631,6 |
| **Total Time** | 1199,3 | 100 | 83,4 | 1961,2 | 100 | 51 |

As can be seen in Table 3, the implementation based on lookup tables provides better time and throughput results in both platforms. In particular, it can be considered the fastest procedure for multiplication because it results in an improvement higher than 90% in the total time consumption with respect to the recursive method in iOS. According to the results obtained for Android implementation, it presents an improvement of around 70% in total time consumption. Thus, in both cases, the results show that the most efficient method for multiplication implementation is without a shadow of a doubt the lookup table. On the other hand, the data obtained for iOS platform show an improvement of around 40% in the required time for shifting the LFSR while Android platform presents an improvement of around 10%.

However, one of the biggest problems with this proposal based on lookup tables might be the need of storage in devices with limited resources. In particular, for SNOW 3G, the required table has 256 elements, each of 32 bits, what results in a total of 32*256 bits. Furthermore, the implementation uses two functions MULα and DIVα, which involve two tables, meaning a total of 2048 bytes. Consequently, this method could be adequate for the characteristics of the chosen devices because both are smartphones with enough storage capacity.

## 4.2 LFSR

Traditionally, the LFSRs used in many stream ciphers are defined over the binary Galois Field GF(2). However, in some specific cases, such as SNOW 3G, the used registers work on extension fields in order to take advantage of the underlying structure of current processors. Besides, the LFSR structures are difficult to implement efficiently in software. The main reason of this fact is the shift of each position during each clock pulse. This shift in hardware implementation occurs simultaneously, so the whole process can be performed in a single clock pulse. However, in software implementation, the process is iterative and costly.

As we saw in Table 3, once the multiplication is optimized, the ClockLFSRKeyStreamMode function is the most time consuming. Hence, this section aim is to implement different LFSR shift techniques for SNOW 3G proposed in [17] in order to determine whether it is possible to minimize run times. Apart from the different software optimization techniques proposed in [17], we also implemented the hardcode technique presented in the SNOW 3G specifications in order to analyse whether some of these proposals can be used to improve the LFSR's final performance.

### Hardcode

The hardcode method consists in embedding the data directly into the source code, instead of using loops or indices, as the rest of techniques do. The cost of this proposal corresponds to 15 assignments. This technique, despite being longer, seems to require less time. Below we can find the implementation of this method.

```
void ClockLFSRKeyStreamMode(){
u32 v = ( ( (LFSR_S0<<8) &0xffffff00 ) ^
 ( MULalpha( (u8)((LFSR_S0>>24) &0xff) ) ) ^
 ( LFSR_S2 ) ^
 ( (LFSR_S11>>8) &0x00ffffff ) ^
```

```
 ( DIValpha( (u8)( ( LFSR_S11) &0xff ) ) ));
LFSR_S0 = LFSR_S1;
LFSR_S1 = LFSR_S2;
LFSR_S2 = LFSR_S3;
LFSR_S3 = LFSR_S4;
LFSR_S4 = LFSR_S5;
LFSR_S5 = LFSR_S6;
LFSR_S6 = LFSR_S7;
LFSR_S7 = LFSR_S8;
LFSR_S8 = LFSR_S9;
LFSR_S9 = LFSR_S10;
LFSR_S10 = LFSR_S11;
LFSR_S11 = LFSR_S12;
LFSR_S12 = LFSR_S13;
LFSR_S13 = LFSR_S14;
LFSR_S14 = LFSR_S15;
LFSR_S15 = v;}
```

### Traditional Approach

The software implementation of the register is essentially based on an array containing 16 integer values of type int. The shift of the register is implemented as a loop that is carried out repeatedly LFSR length times. In each loop iteration, the position *i* is replaced for the contiguous position *i+1*.

### Circular Buffer

This method consists on moving several pointers along the register instead of moving its content. When the pointers reach the end of the register (buffer), they return to the beginning as a circular buffer. Updating the different indexes involves using modular arithmetic, where the module is the length of the register L. Besides, if L is a power of 2, this operation is very efficient in current processors. This technique avoids the expensive L-1 shift in a register of L stages required for the traditional method.

Fig. 2 shows how the pointers update their positions for the different iterations. The pointer *PI* always indicates the position of the *s0* stage, that is to say, the beginning of the register. On the other hand, *PF* indicates the *S15* position and hence the end of the register.

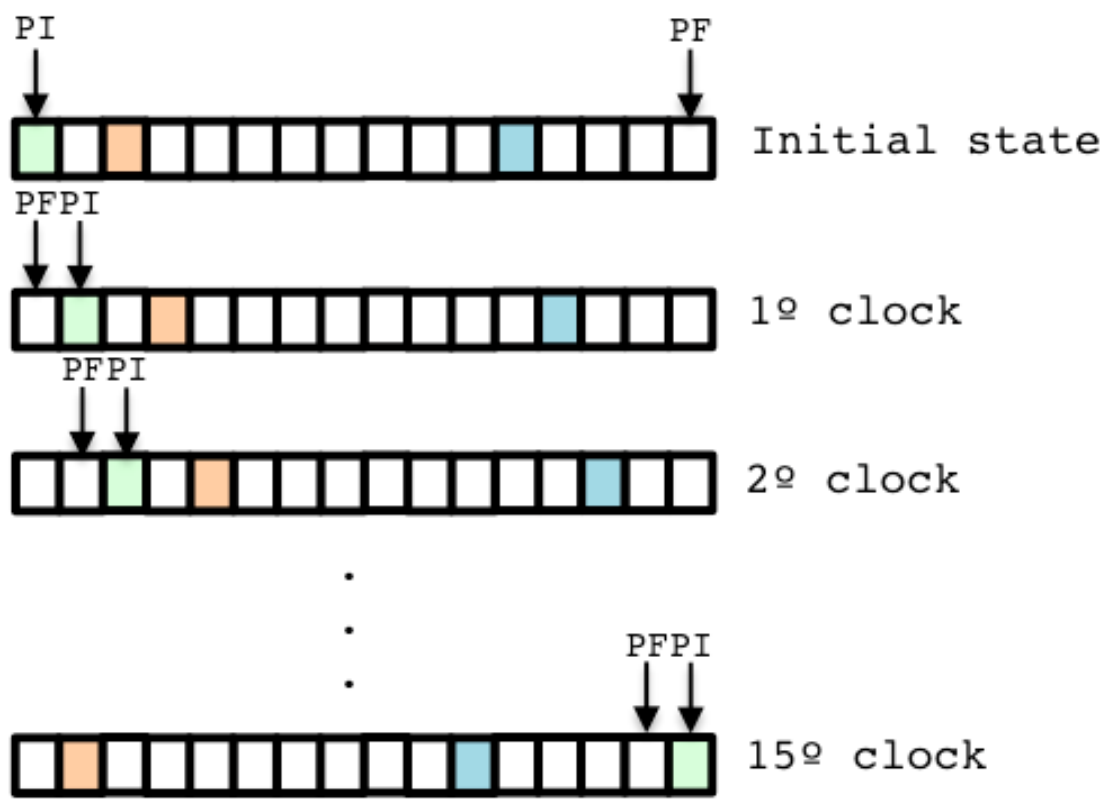

**Fig. 2.** Pointers Movement for Circular Buffer Method

Another important aspect to consider is the position of the different stages that are used in the feedback polynomial. These positions change as pointers move along the register. These positions are highlighted in Fig. 2 in green, orange and blue, corresponding to stages s0, s2 and s11, respectively. In order to store these positions, the code implements an array, which is updated as the pointers PF and PI that are moved along the register. These positions are also taken module L.

### Sliding Windows

Sliding window is a software mechanism for flow control that solves two major problems in TCP/IP data such as flow control and efficiency in transmission. Furthermore, this technique is used in the Authentication Header (AH) protocol, which is applied to provide packet-level security for both IPv4 and IPv6. In particular, this protocol AH uses sliding windows against replay attacks.

This proposal is based on the idea of sliding window to implement the LFSR. In this case, the register size is twice its normal size and its content is duplicated in the second half (see Fig. 3). This proposal only implements a pointer that is initialized to the second half of the register, which indicates the register ending position for each iteration. When the pointer reaches the register end, it returns to its initial position.

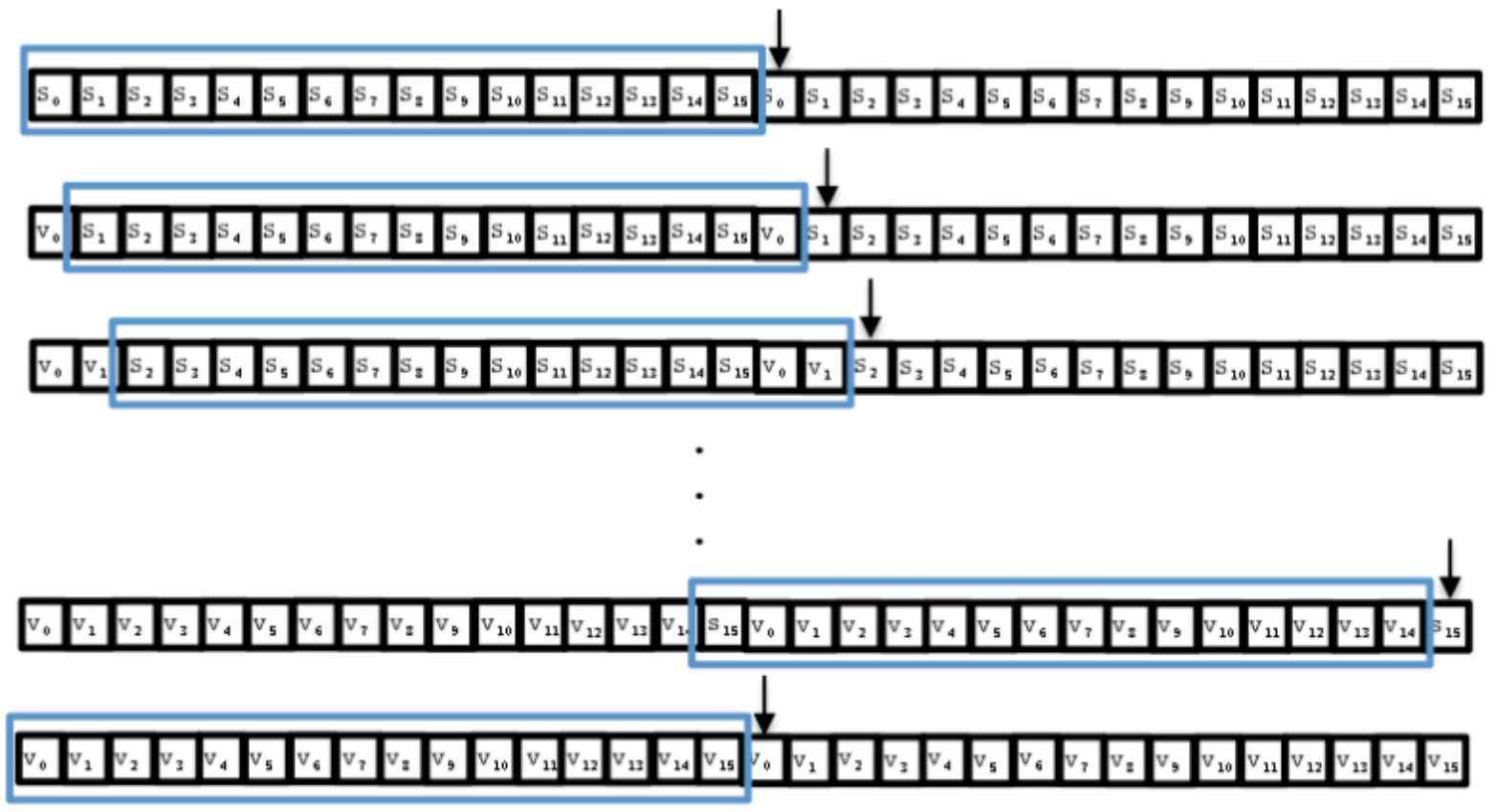

**Fig. 3.** Windows Movement for Sliding Windows Method

When a new value is calculated to be included in the register, it will be written in two different places: in the first position of the window and in the position indicated by the pointer. Accesses to these register positions are fixed according to the current pointer position for each clock. For this reason, the time required to update register information remains constant, which makes this proposal very interesting for its use.

## Loop Unrolling

This mechanism is a basic compiler optimization that reduces the loop overhead by reducing the number of iterations and replicating the body of the loop. It is a quite straightforward technique and very powerful.

This technique consists of repeating several iterations of a loop within the loop body and adjusting the loop index accordingly, by improving its speed at the expense of increasing the program size. The loop unrolling purpose increases program speed to reduce the instructions that control the loop, such as pointer arithmetic or final loop verification for each loop iteration. In order to eliminate this overhead in computation, the loops can be rewritten as a repetition of similar statements.

## Final Results

The analysis carried out with the multiplication method based on lookup tables involves an experiment to assess $10^7$ bytes of the keystream generated by the LFSR of the SNOW 3G. The result values for *ClockLFSRKeyStreamMode* are summarized in Table 4, which shows the total implementation time for each proposed method.

**Table 4.** Different LFSR Implementation Performance for *ClockLFSRKeyStreamMode*

| ***Traditional*** | | ***HardCode*** | | ***Circular Buffers*** | | ***Sliding Windows*** | | ***Loop Unrolling*** | |
|---|---|---|---|---|---|---|---|---|---|
| ***Time (ms)*** | | ***Time (ms)*** | | ***Time (ms)*** | | ***Time (ms)*** | | ***Time (ms)*** | |
| ***iOS*** | ***Android*** | ***iOS*** | ***Android*** | ***iOS*** | ***Android*** | ***iOS*** | ***Android*** | ***iOS*** | ***Android*** |
| 491,1 | 1772,05 | 342,5 | 1513,97 | 834 | 1057,719 | **184,1** | **943,569** | 291,3 | 1697,843 |

The results of Table 4 show that the hardcode method is not the best implementation. Although it represents a 30,26% of improvement over the traditional method for iOS platform and a 14,5% for Android platform, the sliding windows method presents an improvement of 62,5% for iOS and 89,6% for Android with respect to the traditional method. Furthermore, results clearly show the difference between the two types of languages used. For the iOS platform, sliding windows represent an improvement of 46,25% of improvement over the proposed hardcode while Android platform has a difference of 87,84%. This is because Java handles implementation of sliding window much better. Anyway, it is clear that in both cases this is the best proposal. According to the results, we can conclude that regardless of the used architecture and language, the best implementation for the LFSR implementation is with sliding windows.

It can be thought that different proposals for optimizing the implementation of the LFSR might affect other SNOW 3G parts like the FSM. For this reason in our final experiment we tried to determine whether improving LFSR shift time affects negatively other code parts or not. Besides, we analysed the total improvement that can be achieved by combining different methods. In order to do it, for instance we implemented SNOW 3G with lookup tables using sliding windows for the LFSR, and Table 5 shows the summary of the obtained results.

If we compare them with the results of Table 3, it is clear that iOS implementation improves ClockLFSRKeyStreamMode and S1 functions. However, other functions like ClockFSM, GenerateKeystream, S2 or main have increased slightly their times. In fact, the function with the worst time result is S2 as its value has increased 26% with respect to the previous proposal. On the other hand, the greatest improvement, of 47%, happened in the ClockLFSRKeyStreamMode function. Similar results were obtained in the Android platform, where the obtained improvement related to the ClockLFSRKeyStreamMode function was of 80% while the rest of the functions increased their times. In conclusion, the optimized implementations produce an overall improvement in both platforms of about 10% with respect to the code described in the specifications.

**Table 5.** Function Performance in Optimized Mode

| **Computational Cost** | | | | | | |
|---|---|---|---|---|---|---|
| | **iOS** | | | **Android** | | |
| **Function** | ***Time (ms)*** | % | ***Throughput (Mbps)*** | ***Time (ms)*** | % | ***Throughput (Mbps)*** |
| ClockLFSRKeyStreamMode | 184,7 | 17,1 | 541,4 | 261,5 | 14,8 | 382,4 |
| main | 282,3 | 26,1 | 354,2 | 618,6 | 34,9 | 161,7 |
| ClockFSM | 195,4 | 18,1 | 511,8 | 427,8 | 24,1 | 233,8 |
| S1 | 135,9 | 12,6 | 735,8 | 65 | 3,7 | 1538,5 |
| S2 | 163,4 | 15,1 | 612 | 65,8 | 3,7 | 1519,8 |
| GenerateKeystream | 118,2 | 10,9 | 846 | 331,2 | 18,7 | 301,9 |
| Generator | 1,2 | 0,1 | 83333,3 | 2,1 | 0,1 | 47619 |
| **Total Time** | 1081,1 | 100 | 92,5 | 1772 | 100 | 56,4 |

## 5 Conclusions and Future Work

This paper has provided an analysis from a practical point of view of the generator used for the protection of confidentiality and integrity in the 4G/LTE generation of mobile phones. In particular, several software implementations of the SNOW 3G generator that forms the basis of 4G/LTE have been done, both on the iOS and the Android mobile platforms. Thus, after performing several experiments, and comparing the obtained results with others obtained with similar software, several interesting conclusions on how to improve efficiency through the optimization of the software were deduced. Specifically, this work has provided a study of different solutions to address both the implementation problem of both multiplication and LFSR operations in the SNOW 3G generator. In order to determine which are the main reasons that worsen software implementations of the LFSR and/or the FSM in the SNOW 3G generator, and especially with the intention to propose an optimized solution for the problem, different approaches were designed and combined to produce improved results. In particular, the minimization of the total time required for the SNOW 3G generator has been possible by combining the method of lookup tables for the multiplication and of sliding windows for the LFSR.

A possible future work is the implementation of the SNOW 3G generator with other proposals to improve different parts, in order to check whether more improvements can be achieved. Also it would be interesting to validate the proposal of a lightweight version of the SNOW 3G generator for devices with limited resources such as wearable devices, and to analyze thoroughly some theoretical properties of the generator that could be used to improve its implementation.

## Acknowledgements

Research supported by Spanish MINECO and European FEDER Funds under projects TIN2011-25452, IPT-2012-0585-370000 and RTC-2014-1648-8.

## References

1. ETSI/SAGE Specification of the 3GPP Confidentiality and Integrity Algorithms UEA2 & UIA2. Document 2: SNOW 3G Specification, version 1.1, 2006.
2. Ekdahl, P., Johansson, T.: SNOW - a new stream cipher, First Open NESSIE Workshop, pp. 167-168, 2000.
3. Ekdahl, P., Johansson, T.: A New Version of the Stream Cipher SNOW. Selected Areas in Cryptography, Lecture Notes in Computer Science 2595, pp. 47–61, 2003.
4. Hawkes, P., Rose, G.G.: Guess-and-determine attacks on SNOW. Selected Areas in Cryptography, Lecture Notes in Computer Science 2595, pp. 37-46, 2003
5. Coppersmith, D., Halevi, S., Jutla, C.: Cryptanalysis of stream ciphers with linear masking, CRYPTO, Lecture Notes in Computer Science 2442, pp. 515-532, 2002.
6. Berbain, C., Billet, O., Canteaut, A., Courtois, N., Gilbert, H., Gouget, A., Sibert, H.: Sosemanuk, a fast software-oriented stream cipher. eSTREAM, ECRYPT Stream Cipher Project, 2006.

7. ISO/IEC 18033-4: Information technology - Security techniques - Encryption algorithms - Part 4: Stream ciphers, 2005.
8. Billet, O., Gilbert, H.: Resistance of SNOW 2.0 Against Algebraic Attacks. Topics in Cryptology CT-RSA, Lecture Notes in Computer Science 3376, pp. 19–28, 2005.
9. ETSI/SAGE: Specification of the 3GPP Confidentiality and Integrity Algorithms UEA2 & UIA2. Document 5: Design and Evaluation Report, Version 1.1, 2006.
10. Nyberg, K., Wall´en, J.: Improved Linear Distinguishers for SNOW 2.0. Fast Software Encryption FSE, Lecture Notes in Computer Science 4047, pp. 144–162, 2006.
11. Watanabe, D., Biryukov, A., De Canniere, C.: A Distinguishing Attack of SNOW 2.0 with Linear Masking Method. Selected Areas in Cryptography SAC, Lecture Notes in Computer Science 3006, pp. 222–233, 2004.
12. Debraize, B., Corbella, I.M.: Fault Analysis of the Stream Cipher Snow 3G, Workshop on Fault Diagnosis and Tolerance in Cryptography, pp. 103–110, 2009.
13. Brumley, B., Hakala, R., Nyberg, K., Sovio, S.: Consecutive S-box Lookups: A Timing Attack on SNOW 3G, Information and Communications Security, Lecture Notes in Computer Science 6476, pp. 171–185, 2010.
14. Biryukov, A., Priemuth-Schmid, D., Zhang, B.: Multiset collision attacks on reduced-round SNOW 3G and SNOW 3G. Applied Cryptography and Network Security, pp. 139-153, 2010.
15. Orhanou, G., El Hajji, S., Bentaleb, Y.: SNOW 3G stream cipher operation and complexity study. Contemporary Engineering Sciences-Hikari Ltd, 3(3), pp. 97-111, 2010.
16. Kitsos, P., Selimis, G., Koufopavlou, O.: High performance ASIC implementation of the SNOW 3G stream cipher, IFIP/IEEE VLSI-SOC, pp. 13-15, 2008.
17. Delgado-Mohatar, O., Fúster-Sabater, A.: Software Implementation of Linear Feedback Shift Registers over Extended Fields, CISIS/ICEUTE/SOCO, pp. 117-126, 2012.
18. Molina-Gil, J., Caballero-Gil, P., Caballero-Gil, C., Fúster-Sabater, A.: Analysis and implementation of the SNOW 3G generator used in 4G/LTE systems. CISIS/ICEUTE/SOCO, pp. 499-508, 2014.
19. Jenkins, C., Schulte, M., Glossner, J.: Instructions and hardware designs for accelerating SNOW 3G on a software-defined radio platform, Analog Integrated Circuits and Signal Processing 69 (2-3), pp. 207-218, 2011.